\documentclass[numberedappendix,twocolumn,tighten]{aastex62}

\usepackage{graphicx, hanging}
\usepackage[caption=false]{subfig}
\begin{document}

\title{Stellar and Molecular Gas Rotation in a Recently-Quenched Massive Galaxy at \lowercase{z}~$\sim$~0.7}
\author{Qiana Hunt}
\affiliation{Department of Astrophysics, Princeton University, Princeton, NJ 08544, USA}
\author[0000-0001-5063-8254]{Rachel Bezanson}
\affiliation{Department of Physics and Astronomy and PITT PACC, University of Pittsburgh, Pittsburgh, PA, 15260, USA}
\author{Jenny E. Greene}
\affiliation{Department of Astrophysics, Princeton University, Princeton, NJ 08544, USA}
\author[0000-0003-3256-5615]{Justin S. Spilker}
\affiliation{Department of Astronomy, University of Texas at Austin, 2515 Speedway, Stop C1400, Austin, TX 78712, USA}
\author[0000-0002-1714-1905]{Katherine A. Suess}
\affiliation{Astronomy Department, University of California, Berkeley, CA 94720, USA}
\author{Mariska Kriek}
\affiliation{Astronomy Department, University of California, Berkeley, CA 94720, USA}
\author[0000-0002-7064-4309]{Desika Narayanan}
\affiliation{Department of Astronomy, University of Florida, 211 Bryant Space Sciences Center, Gainesville, FL 32611, USA}
\affiliation{University of Florida Informatics Institute, 432 Newell Drive, Gainesville, FL 32511, USA}
\affiliation{Cosmic Dawn Center (DAWN), Niels Bohr Institute, University of Copenhagen, Juliane Maries vej 30, DK-2100 Copenhagen, Denmark}
\author[0000-0002-1109-1919]{Robert Feldmann}
\affiliation{Institute for Computational Science, University of Zurich, CH-8057 Zurich, Switzerland}
\author{Arjen van der Wel}
\affiliation{Sterrenkundig Observatorium, Universiteit Gent, Krijgslaan 281 S9, B-9000 Gent, Belgium}
\affiliation{Max-Planck Institut f{\"u}r Astronomie, K{\"o}nigstuhl 17, D-69117, Heidelberg, Germany}
\author{Petchara Pattarakijwanich}
\affiliation{Department of Physics, Faculty of Science, Mahidol University, Bangkok 10400, Thailand}

\begin{abstract}
The process by which massive galaxies transition from blue, star-forming disks into red, quiescent galaxies remains one of the most poorly-understood aspects of galaxy evolution. In this investigation, we attempt to gain a better understanding of how star formation is quenched by focusing on a massive post-starburst galaxy at z = 0.747. The target has a high stellar mass and a molecular gas fraction of ~30\% | unusually high for its low star formation rate. We look for indicators of star formation suppression mechanisms in the stellar kinematics and age distribution of the galaxy obtained from spatially resolved Gemini Integral-Field spectra and in the gas kinematics obtained from ALMA. We find evidence of significant rotation in the stars, but we do not detect a stellar age gradient within 5 kpc. The molecular gas is aligned with the stellar component, and we see no evidence of strong gas outflows. Our target may represent the product of a merger-induced starburst or of morphological quenching; however, our results are not completely consistent with any of the prominent quenching models.

\end{abstract}

\section{Introduction} \label{intro}

Luminous galaxies in the low- and intermediate-redshift universe can generally be divided into two distinct populations: blue, active spirals; and red, quiescent ellipticals \citep[e.g.][]{strateva01, alatalo14}. Red galaxies, which harbor little or no star formation, must have formed their stars at some point in the early universe and thus are the descendants of once-active blue galaxies \citep[e.g.][]{bell04, faber07}. The transition from one stage to the other requires the suppression, or quenching, of star formation. Little is known about the specific processes responsible for quenching massive galaxies. Recent studies suggest a strong correlation between structural changes in the galaxy and star formation suppression \citep[e.g.][]{newman15, yano16}. Several models have been proposed. For instance, \textit{gas-poor mergers} may suppress star formation by heating the surrounding gas halo via inflow-triggered shocks \citep{johansson09, naab09, hopkins09}. \textit{Gas-rich major mergers} may also heat up the gas supply and guard it against gravitational collapse \citep{hopkins08}, or disrupt the cold gas disk. Mergers may lead to \textit{compaction}, in which the star-forming gas migrates inward, triggering a central starburst and inside-out gas depletion \citep{wellons15, tacchella15, zolotov15}, or can trigger high-velocity \textit{outflows} driven by starburst radiation or quasar activity \citep{tremonti07, sell14, pontzen17, mao18}.
Alternatively, galaxies may quench without removing cold gas if the gas is stabilized against clumping by dynamical effects following \textit{gas-rich minor mergers} \citep{vdVoort18} or \textit{morphological quenching} \citep{martig09}.

Different quenching scenarios should leave characteristic marks on the distribution of ages and motions of stars within quiescent galaxies.
However, these signatures gradually diminish over time \citep{cananzi93, goto03, kauff03}. In order to better understand the quenching process, it is necessary to observe galaxies soon after they transition. Post-starburst galaxies (PSBs) -- galaxies that suddenly turned quiescent after a period of intense star formation -- are ideal subjects. Their spectra are dominated by A-type stars with lifespans on the order of $\sim$1 Gyr \citep[e.g.][]{pogg97, leborgne06}, so a lack of current star formation suggests they must have quenched within a few Gyrs before observation. As `transitional objects,' PSBs may represent a bridge between massive blue and red galaxies \citep[e.g.][]{alatalo14}.

\begin{figure*}[t]
\centering
\captionsetup{belowskip=10pt}
    \subfloat[Flux density of SDSS J0912+1523]{
         \includegraphics[width=0.4\textwidth]{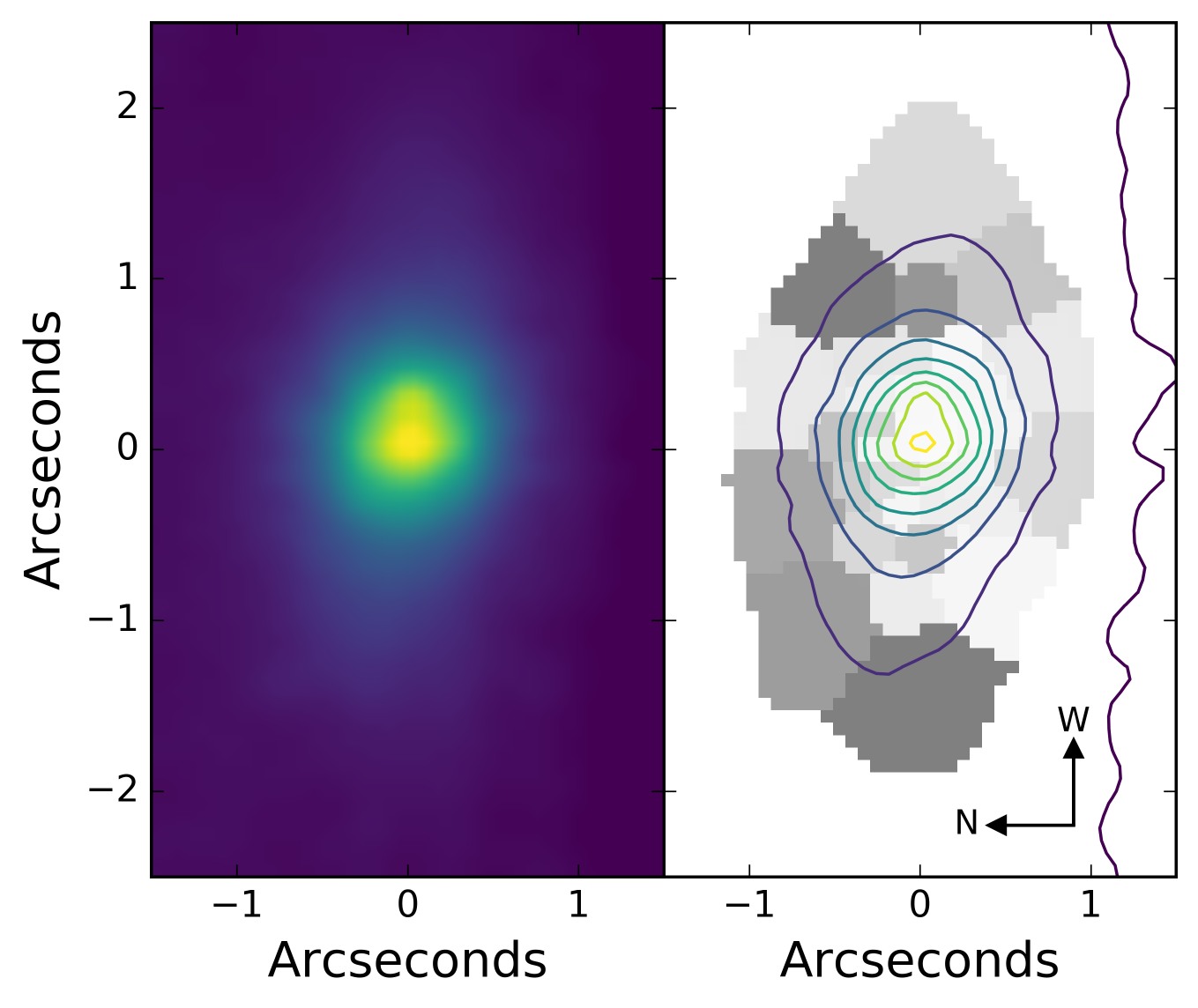}
    }
    \subfloat[Central spectrum and H$\delta_{A}$ bandpasses]{
         \includegraphics[width=0.53\textwidth]{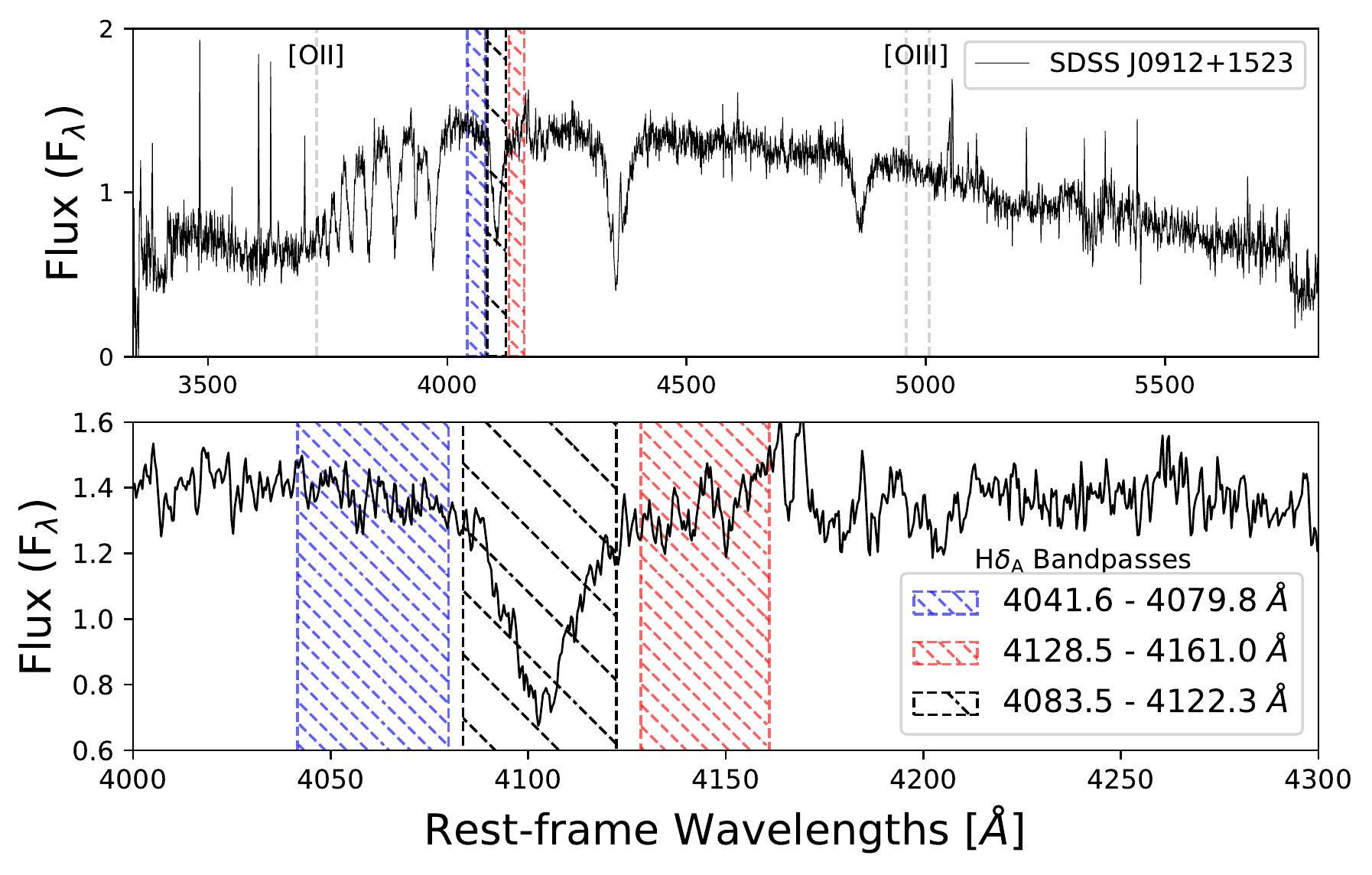}
    }
    \caption{(a) Median flux map of SDSS J0912+1523 from the Gemini spectra, with contours overlaid onto the Voronoi bins. (b) Full spectrum of the central Voronoi bin and the H$\delta_{A}$ central bandpass, surrounded by blue and red `continuum' bandpasses, as defined by \citet{worthey97}. The wavelengths of the [OII]$\lambda$3727, [OIII]$\lambda$4959, and [OIII]$\lambda$5007 lines are labeled.
    \label{indices1}}
\end{figure*} 


In this Letter, we examine an intermediate-redshift PSB, SDSS J0912+1523, with a stellar mass of $\sim$~2$\times$10$\rm^{11}~M_{\odot}$. The target was chosen out of a large sample of PSBs \citep{suess17} selected from the SDSS DR12 catalog \citep{alam15} and included in \citet{pete16}. The galaxies in the PSB sample were identified by their strong Balmer breaks and blue slopes redward of the break, as demonstrated in \citet{kriek10}. SDSS J0912+1523 was chosen as the brightest, most A-star dominated source at the high end of the sample's redshift range, with z~=~0.747 and i$_{AB}$~=~18.6 mag. SDSS J0912+1523 represents a rare opportunity to study the spatially resolved kinematics of a massive, recently-quenched PSB. \citet{suess17} present ALMA observations of the galaxy's molecular gas. 
Here, we analyze the stellar component observed by Gemini to obtain the kinematic properties and age distribution of the stellar population. Combining the stellar and gas data, we identify markers left by the quenching process and constrain the means by which SDSS J0912+1523 suppressed its star formation.
We assume a cosmology of $\Omega_{m}$~=~0.3, $\Omega_{\Lambda}$~=~0.7, and \textit{h}~=~0.7.

\section{Data} \label{method}
We obtained a spatially resolved spectral datacube from observations of SDSS J0912+1523 made with the Gemini North 8m telescope in March 2016. The target was observed using GMOS-N in one-slit IFU mode with the R400 grating, giving a field of view of approximately 7$\arcsec\times$5$\arcsec$. We acquired 5 exposures of 24.5 minutes each. The observations have a spectral FWHM of $\sim$1.35 \AA ~under $\sim$0.5$\arcsec$ seeing conditions, as reported by the Gemini staff. Our data yield spectral coverage over 3343-5820~\AA ~in the rest frame. We reduced the data with \texttt{gemtools} and \texttt{gmos} in the Gemini \texttt{IRAF} package\footnote{http://www.gemini.edu/sciops/instruments/gmos/} using the procedure described by \citet{lena14}.

We collapse our datacube in the spectral dimension by taking the median flux of each spaxel. The resulting 2D flux distribution and contour maps are shown in Figure \ref{indices1}a. We note an asymmetry in the object's core, which may indicate the presence of two peaks. The flux of the secondary peak, $\sim$0.2$\arcsec$ ~west of the galaxy's center, reaches 91\% of the maximum. We investigate the nature of the peaks in Section \ref{results}. 

We remove all spaxels with individual signal-to-noise ratios (S/N) beneath a threshold of 0.3 from the flux map and spatially bin the remainder using an adaptive Voronoi method \citep{capp03}. We set the target S/N of the spatial bins to 6. The median galaxy spectrum of the spaxels in each bin is then calculated and examined individually.

\subsection{Spatially Resolved Kinematics} \label{using_ppxf}
\begin{figure*}[ht]
\centering
\includegraphics[width=\textwidth]{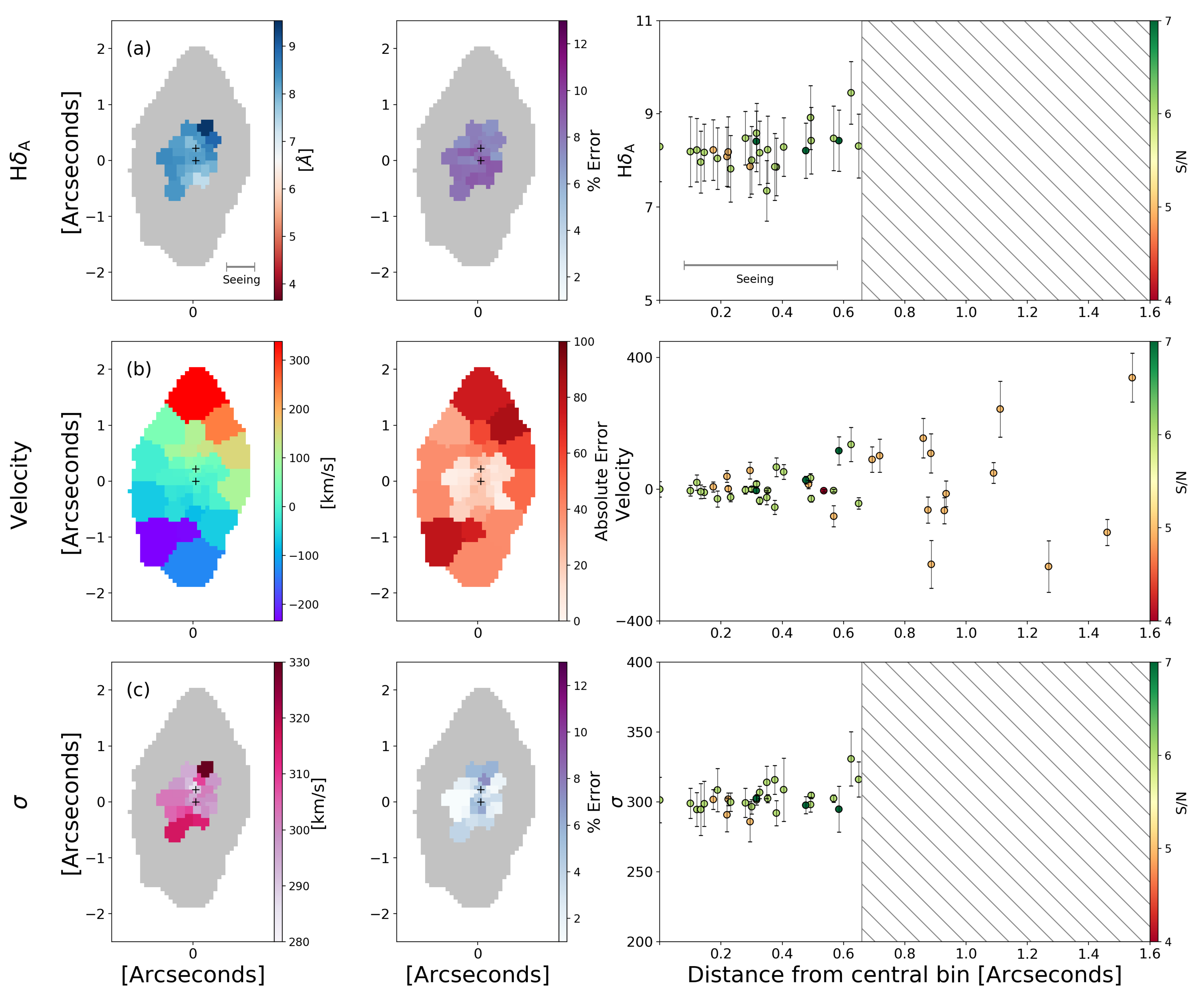}
\captionsetup{belowskip=6pt}
\figcaption{Stellar kinematic and spectral index maps. The two flux peaks are marked with black + symbols. The seeing is indicated with grey bars. The left column shows the measured values of the (a) equivalent width of H$\delta_{A}$, (b) velocity, and (c) velocity dispersion $\sigma$ for each Voronoi bin. The middle column represents the estimated errors for each measurement. Velocity (row b) is measured with respect to the central bin, and its errors are presented as absolute values. The right column illustrates the relationship between the measured values and radial distance from the center of the galaxy.  We omit outer bins below the target S/N = 6 in rows (a) and (c). These bins are blocked out in grey in the left column and by grey hashed boxes in the right column.
\label{J0912}}
\end{figure*}

The Penalized Pixel-Fitting (pPXF) software \citep{capp04, capp17} is used to extract stellar kinematics from each Voronoi bin by fitting a set of higher-resolution template spectra to the galaxy spectrum. We use theoretical simple stellar population model templates with a resolution of R~=~10,000 created by Charlie Conroy (Bezanson et al. in prep). We assume solar metallicity and mask the strong telluric features in the galaxy spectrum. The template and median spectra are fed into pPXF, which convolves the templates to the observed spectral resolution and fits a non-negative combination of the templates to the spectrum, along with a multiplicative and additive polynomial that accounts for dust reddening in addition to potential issues with flux calibration. The code returns a best fitting stellar population model, stellar velocity, and velocity dispersion (hereafter referred to as $\sigma$). Maps of velocity and $\sigma$ are shown in Figure ~\ref{J0912}b and c.

The uncertainties in velocity and $\sigma$ are derived using 100 iterations of Monte Carlo resampling, in which the residuals are randomly sampled, added to the best fitting model, and rerun through pPXF as the new ``galaxy" spectrum. The typical uncertainty is 22~km~s$^{-1}$ for stellar velocity and 10~km~s$^{-1}$ for $\sigma$.

\subsection{Measuring Age-Sensitive Spectral Indices} \label{measuring_indices}

The hydrogen Balmer lines are strong absorption features that dominate the spectra of A-type stars. Since they are sensitive to emission-filling by ongoing star formation, the strong presence of these features in the absence of emission lines indicates a burst of star formation followed by sudden suppression and passive evolution \citep[e.g.][]{cananzi93, goto03, kauff03}. For PSB populations, stronger Balmer absorption features signify more recent bursts \citep{kauff03}.
 
We use the equivalent width (EW) of the H$\delta$ Balmer line as a proxy for stellar ages. H$\delta$ is well-isolated from other spectral features and less susceptible to contamination from emission-filling than other Balmer lines \citep{goto03}. We measure H$\delta_{A}$ using the wide \citet{worthey97} bandpass definition, illustrated in Figure ~\ref{indices1}b, with the \texttt{lick\_ew} routine from the IDL EZ\_Ages package written by G. Graves \citep{schiavon07, graves08}. The uncertainties are calculated via Monte Carlo resampling, as described in Section \ref{using_ppxf}. The typical uncertainty in H$\delta_{A}$ is $\sim$0.7 \AA. 

\begin{figure*}[ht]
\centering
\includegraphics[scale=.55]{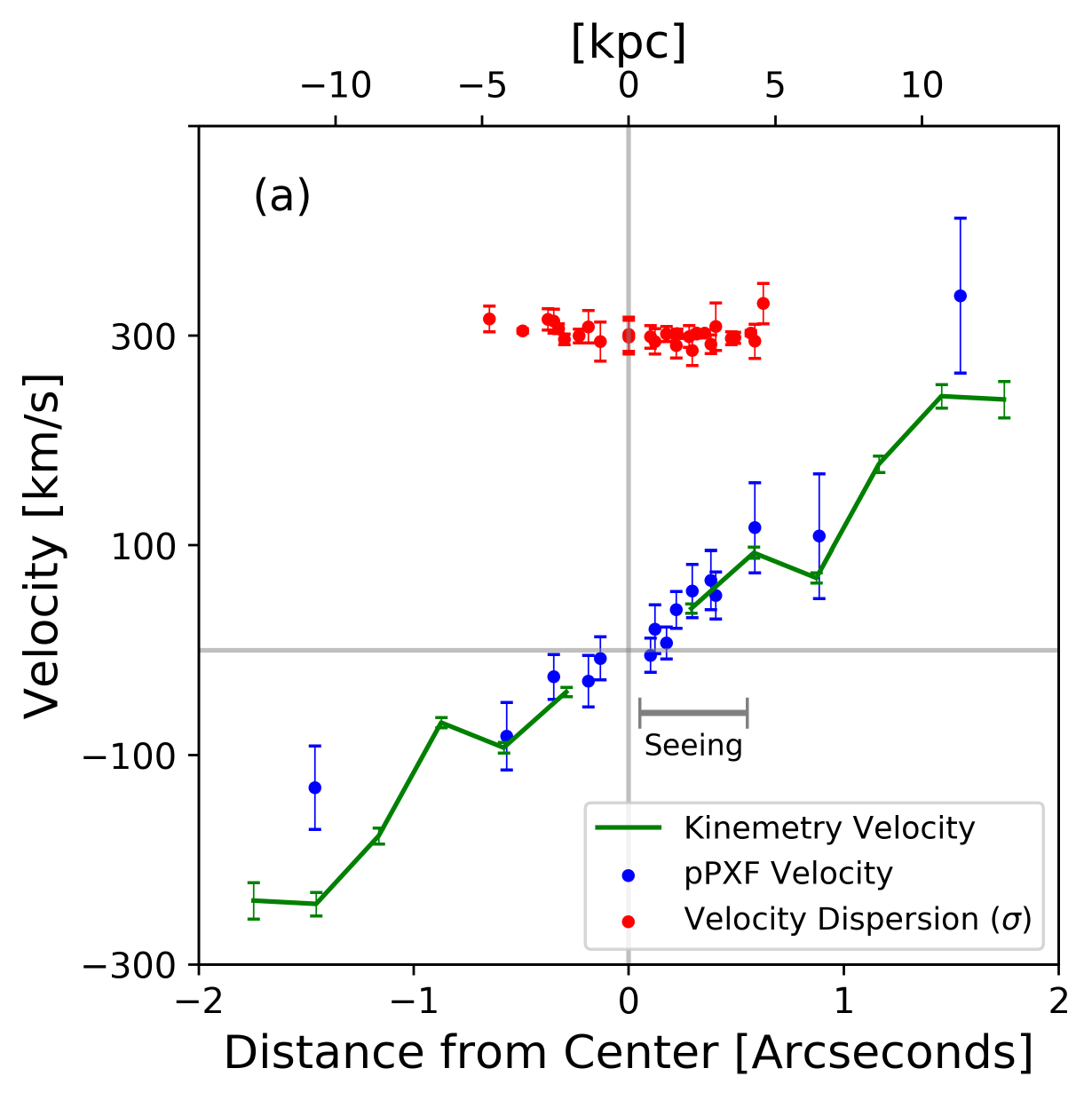}
\hspace{-.3cm}
\includegraphics[scale=.54]{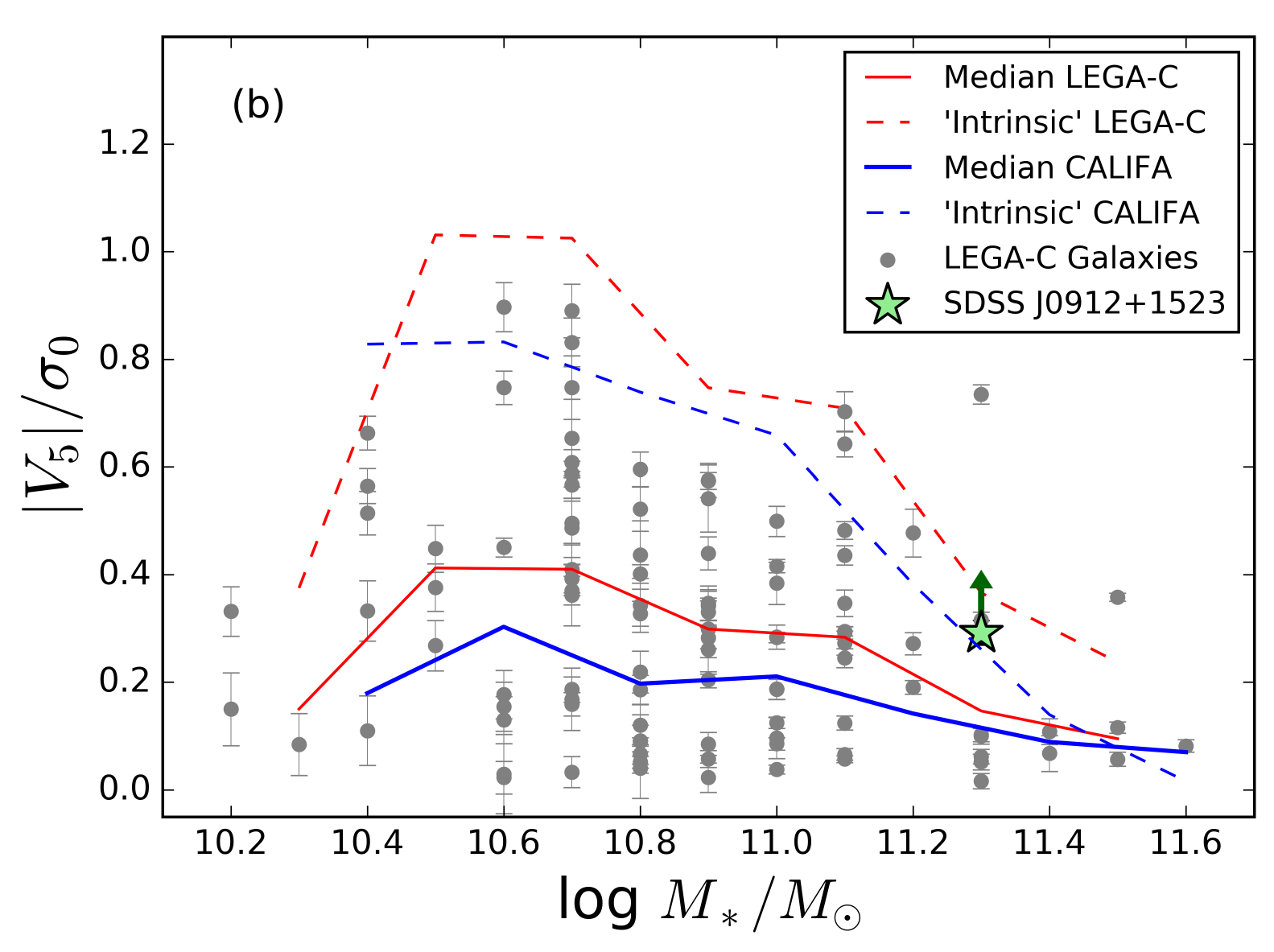}
\captionsetup{belowskip=10pt}
\figcaption{(a) Velocity and velocity dispersion of the SDSS J0912+1523 Voronoi bins as a function of distance from the center. (b) Rotational support at 5 kpc vs. stellar mass of SDSS J0912+1523, compared to similar-redshift quiescent galaxies from the LEGA-C survey and local galaxies from CALIFA \citep{bez2018}. The median observed rotational support of the LEGA-C and CALIFA galaxies per mass bin are plotted in red and blue respectively, with dashed lines reflecting intrinsic values. The intrinsic rotational support of the LEGA-C sample and SDSS J0912+1523 (indicated by the green arrow) are simulated with using the CALIFA sample. SDSS J0912+1523 is roughly consistent with the typical, similar-mass LEGA-C galaxy and is more rotationally-supported than local CALIFA galaxies. \label{lega-c}}
\end{figure*}

The D$_n$4000 index, which measures the strength of the 4000 \AA ~break, is often used in conjunction with H$\delta_{A}$ to probe the star formation history of galaxies \citep[e.g.][]{pogg97, kauff03, kriek11, zahid15, zahid17}. The feature is measured using only two bandpasses \citep{balogh99} and is therefore highly sensitive to the flux calibration. In addition, since D$_n$4000 is especially uncertain in the outer regions of our galaxy due to sky subtraction challenges, we use only the value of the central spatial bin to compare our analysis of SDSS J0912+1523 with previous work.

\section{Results} \label{results}

Figure \ref{J0912} shows the spectral index and stellar kinematics for each Voronoi bin as measured by \texttt{lick\_ew} and pPXF. The left column represents our measurements of (a) H$\delta_{A}$, (b) velocity, and (c) $\sigma$. The colormap limits for  H$\delta_{A}$ were chosen to reflect that of the sample of z~$\sim$~0.6 PSBs discussed in \citet{suess17}. The outer bins below S/N = 6 are omitted from the H$\delta_{A}$ and $\sigma$ maps, as their low S/N makes them unreliable for measuring the EW and broadening of spectral features. We note that there exist a few central bins with S/N~$<$~6 where the Voronoi process has failed to properly expand the bins. As these bins are small and have values consistent with the higher S/N regions, we have chosen to include them in our analysis. The middle column of Figure \ref{J0912} gives the uncertainty of each value. For H$\delta_{A}$ and $\sigma$, the uncertainty is given as a percentage error. Since all velocities are measured with respect to the central bin, the uncertainty in the velocity is given as an absolute error. Finally, the right column shows the radial gradient of each property.

\subsection{Spectral Indices} \label{spect_indices}

We measure a central H$\delta_{A}$ of $\sim$8.3 \AA ~and D$_n$4000 of ~$\sim$1.18. Our measurements are in agreement with the spectral indices obtained from the SDSS data in \citet{suess17}. These values place SDSS J0912+1523 at the boundary between typical starburst and post-starburst galaxy populations \citep{kauff03, suess17}. To ensure SDSS J0912+1523 is indeed quenched, we search for the presence of star formation using [OII]$\lambda$3727 and [OIII]$\lambda$5007. Both emission lines are absent from our spectra, consistent with the low SFR of 2.1 $\pm$ 0.8 M$\rm _{\odot} ~yr^{-1}$ measured previously by \citet{suess17} using the line flux of the [OII]$\lambda$3727 doublet. 
SDSS J0912+1523, therefore, most likely represents a transitional object that has only just quenched. 

H$\delta_{A}$, as illustrated in Figure \ref{J0912}a, appears flat in the inner $\sim$0.6$\arcsec$ region of the galaxy. In addition, we measure H$\delta_{A}$ and D$_n$4000 to be roughly the same for each of the peaks observed in Figure \ref{indices1}a, which are marked with black + symbols in Figure \ref{J0912}. The reported $\sim$0.5$\arcsec$ seeing introduces some uncertainty as to whether we resolve the H$\delta_{A}$ map. Since we are able to resolve a $\sim$0.2$\arcsec$ asymmetry within the core, our seeing is likely a conservative estimate. We believe the flatness of the H$\delta_{A}$ (age) distribution is real, and that the two `cores' we observe in Figure \ref{indices1}a likely share a common stellar population.

\subsection{Kinematics} \label{kin} 

The stellar kinematics of SDSS J0912+1523 are illustrated by rows (b) and (c) in Figure \ref{J0912}, representing stellar velocity and $\sigma$ respectively. Due to beam smearing, the intrinsic velocity will likely be higher than we measure, while the intrinsic $\sigma$ will be lower. The velocity field shows ordered rotation about the center of SDSS J0912+1523, while the $\sigma$ field remains flat. The velocities at the positions of the two peaks are offset by $\sim$40~km~s$^{-1}$, compared to the galaxy's maximum velocity of 338~$\pm$~74~km~s$^{-1}$. Considering the average uncertainty of 22 km s$^{-1}$, the offset is consistent with the velocity gradient with respect to the center of the galaxy. Based on the consistent  
velocity offset, lack of multiple $\sigma$ peaks, and similar stellar populations, we assert that the two `cores' are part of the same galaxy and are rotating as a single object. They may be the remnants of a late-stage merger, or possibly a single core that is bisected by a dust lane. We note that the orientation of such a dust lane is not obviously consistent with the galaxy alignment and orientation.

We quantitatively compare the velocity and $\sigma$ profiles in Figure~\ref{lega-c}a. We measure a smooth velocity field for SDSS J0912+1523 from its 2D line-of-sight velocity distribution using \texttt{Kinemetry} in IDL \citep{kraj06}, represented by the green line. Kinemetry allows us to account for possible changes in B/A as a function of redshift and gives velocities with smaller errors. The maximum velocity is 242~$\pm$~17~km~s$^{-1}$ at $\sim$11 kpc. The pPXF velocities along the galaxy's major axis are marked in blue. Deviations from the smooth Kinemetry model are consistent with the noise.
Using the central velocity dispersion ($\sigma$ of the central bin, $\sigma_{0}$) and the rotational velocity measured from the \texttt{Kinemetry} velocity profile at 5 kpc (V$_{5}$), we calculate rotational support ($\rm|V_{5}|$/$\sigma_{0}$). The 5~kpc physical parameter was chosen to match data from the LEGA-C sample. Galaxies with high $\rm|V_{5}|$/$\sigma_{0}$ tend to be disk-like in morphology. At low $\rm|V_{5}|$/$\sigma_{0}$, galaxies become pressure-supported and exhibit spheroidal morphologies \citep{em07, em11}. For SDSS J0912+1523, we measure $\sigma_{0}$~=~301~$\pm$~16~km~s$^{-1}$ and V$_{5}$~=~88~$\pm$~5~km~s$^{-1}$. Our estimate of rotational support is a lower limit, in part due to beam smearing, intrinsic inclination, and the limited radial range of our measurements. When measured at 10 kpc, V$_{10}$~=~213~$\pm$~11~km~s$^{-1}$. These values indicate a retention of rotational support in the PSB.

We plot the rotational support of SDSS J0912+1523 against its stellar mass, as obtained by \citet{suess17}, and compare them to a sample of quiescent galaxies at 0.6~$<~z~<$~1.0 from the LEGA-C spectroscopic survey \citep{vdWel16} in Figure~\ref{lega-c}b \citep{bez2018}. Like our data, the LEGA-C observations (grey points) are spatially-resolved and are not corrected for inclination or beam smearing.
\citet{bez2018} perform simulations of the LEGA-C observations with 1.0$\arcsec$ seeing using local quiescent galaxies from the CALIFA sample (blue solid line) and find that the measured $\rm|V_{5}|$/$\sigma_{0}$ values for the LEGA-C and CALIFA galaxies
are likely underestimated by a factor of $\sim$2.5. We plot the median $\rm|V_{5}|$/$\sigma_{0}$ of the LEGA-C galaxies per mass bin (red solid line) and adjust it by the underestimation factor to approximate the intrinsic relation for these quiescent galaxies at $z\sim0.8$ (red dashed line). We simulate the effects of beam smearing by a FWHM~$\sim3.5$~kpc (corresponding to $\sim$0.5$\arcsec$ at this redshift) on the CALIFA data cubes and estimate that our measured $\rm|V_{5}|$/$\sigma_{0}$ of SDSS J0912+1523 (green star) will underestimate the intrinsic value by a factor of $\sim$1.3, which we indicate by the green arrow.

\begin{figure}[t]
\centering
\includegraphics[width=1.05\linewidth]{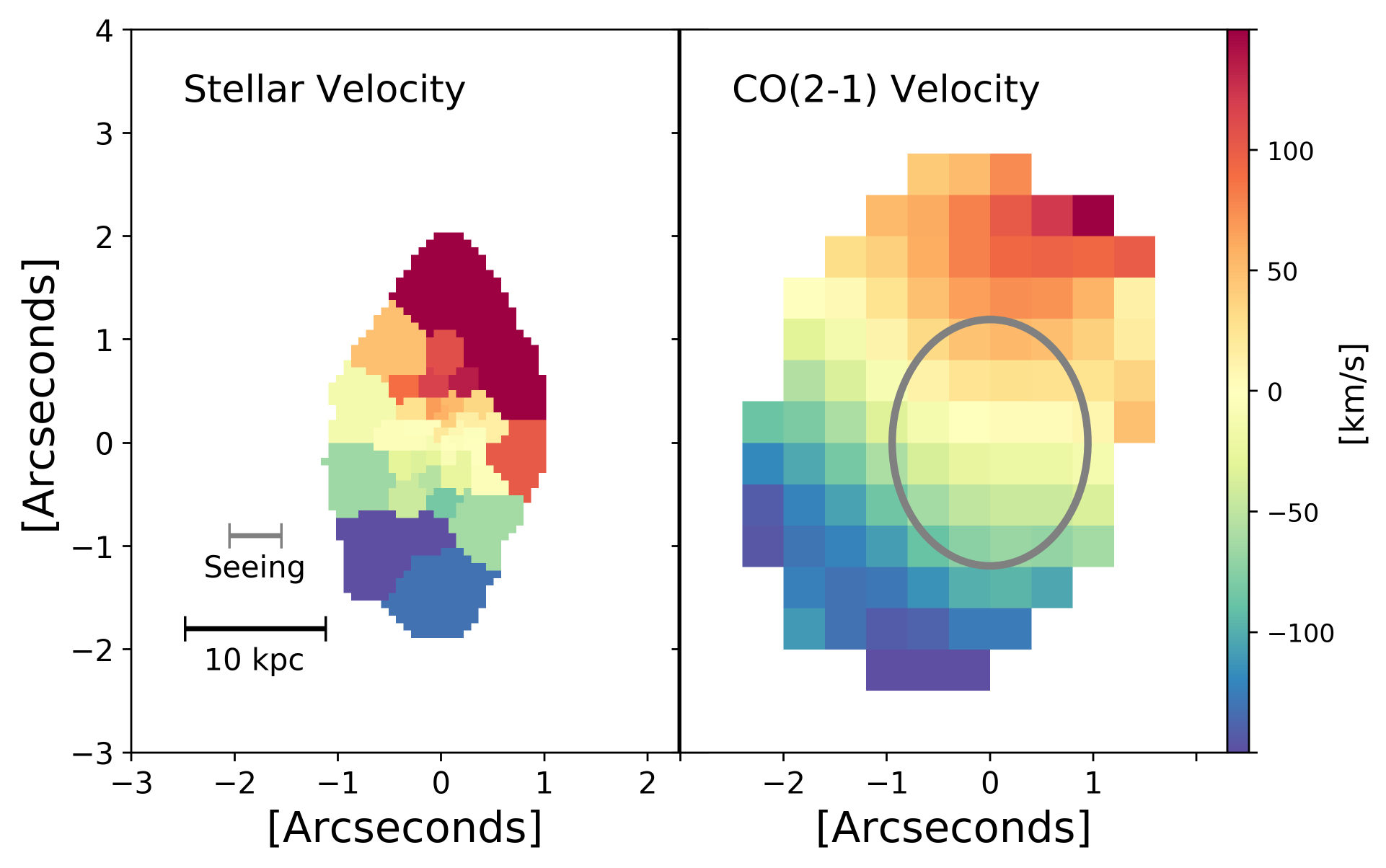}
\figcaption{Stellar and cold molecular gas velocity fields of SDSS J0912+1523, as observed by Gemini and ALMA. The stellar seeing and ALMA beam are indicated in grey. The motion of the cold molecular gas is consistent with the stellar component. \label{alma}}
\end{figure}

The LEGA-C sample reveals a correlation between mass and rotational support, consistent with observations of the nearby universe \citep{em11}. There is a significant increase in pressure-support at log~M$_*/$M$_{\odot}~>$~11.2, above which SDSS J0912+1523 falls. Our target PSB is \textit{at least} consistent with typical quiescent galaxies at similar redshift as probed by LEGA-C, which exhibit stronger rotational support than the nearby CALIFA galaxies of similar mass. We note, however, that beam smearing can only move our galaxy towards higher rotation.

Finally, we compare the stellar and molecular gas components of SDSS J0912+1523. Recent ALMA observations of SDSS J0912+1523 by \citet{suess17} uncovered a substantial amount of cold gas: M$\rm_{gas}$ = (3.4~$\pm$~0.2)~$\times$~10$\rm^{10}~M_{\odot}$, corresponding to a gas fraction of $\sim$30\%. We plot the velocity field  of the molecular gas, as traced by CO(2-1), beside the stellar velocity field in Figure~\ref{alma}. With a spatial resolution of 1.7$\arcsec\times$2.4$\arcsec$, we do not know the spatial distribution of gas. However, based on the similar amplitudes and rotation, we determine the gas and stars are likely aligned.

\section{Discussion and Conclusions} \label{discussion}

Through kinematic and spectroscopic analysis of SDSS J0912+1523 using optical IFU data obtained with Gemini, in conjunction with the analysis of ALMA data first presented in \citet{suess17}, we determine that the massive z $\sim$ 0.7 galaxy:
\begin{itemize}
    \item is a young PSB that has effectively quenched its star formation;
    \item may contain either two cores or a dust lane that bisects a single core;
    \item has no detectable age gradient within 5 kpc of its center, within the constraints of the seeing;
    \item maintains significant rotational support compared to similar-mass quiescent galaxies;
    \item contains an unusually large cold gas fraction for its apparent SFR, and
    \item maintains ordered gas rotation that roughly matches the stellar component, with no evidence of significant outflow.
\end{itemize}

We now compare the characteristics of SDSS J0912+1523 to predictions from the most common quenching scenarios for massive quiescent galaxies. In particular, we focus on: mergers, compaction, outflows, and morphological quenching. 

Based on our observations, none of these scenarios seem to accurately predict all of the characteristics of our target PSB. The significant rotational support and regular velocity field are most consistent with both gas-rich merger and compaction models, while they are inconsistent with the dispersion-dominated systems that result from gas-poor mergers \citep{em07}. 
The abundance of cold molecular gas within SDSS J0912+1523, on the other hand, suggests that neither depletion nor heating of gas led to its quenching, which limits the role of compaction and gas-rich major mergers as the main drivers of star formation suppression. The high gas fraction is more consistent with gas-rich \textit{minor} mergers and morphological quenching. However, it is unclear whether SDSS J0912+1523 contains \textit{too much} gas and rotation for morphological quenching to be viable \citep{martig09}. In addition, gas-rich minor merger models are typically studied at low redshifts, so the galaxies involved in our z~$\sim$~0.7 example likely contained more gas than is normally included in the model \citep{vdVoort18}. The alignment of the gas and stellar velocity fields argue against an external origin for the gas, and there is no evidence for strong molecular gas outflows in the ALMA data. Further constraints arise from the spectral indices and lack of emission lines in the galaxy's spectra. Evidence of a recent starburst is consistent with compaction, though the apparent lack of a radial age gradient may indicate otherwise. It is not clear whether a gas-rich minor merger could fuel a starburst at our redshift.

Whatever process quenched SDSS J0912+1523 must allow it to retain both a large gas fraction and significant rotational support. To definitively constrain the scenarios beyond this, more data are required. Radio or X-ray observations may reveal the presence of quasar activity, while deeper IR and radio observations, as with the VLA, may uncover heavily dust-obscured star formation. Information about the dark matter halo and its environment is also vital, as many quenching models make halo predictions, and the environment in which quiescent galaxies develop may influence their formation histories \citep{zab1996, zolotov15}. \emph{HST} imaging and higher resolution gas observations could greatly improve our understanding of SDSS J0912+1523 by revealing its morphology: whether it is bulge-dominated or disky, how the gas is distributed, whether there are weak outflows, the nature of the `cores', and whether mergers greatly influence the galaxy's behavior. In the meantime, expanding our analysis to include other PSBs may help determine whether SDSS J0912+1523 is a special case, or whether a majority of PSBs -- and possibly quiescent galaxies as a whole -- quench via a similar process.

\begin{acknowledgements}
We wish to thank Charlie Conroy for sharing the stellar population  synthesis model templates with us. 
We also sincerely thank Michael Strauss, Marijn Franx, and Khalil Hall-Hooper for their help, comments, and support.

This paper is based on observations obtained at the Gemini Observatory, acquired through the Gemini Observatory Archive under Program ID GN-2016A-FT-6 and processed using the Gemini IRAF package, which is operated by the Association of Universities for Research in Astronomy, Inc., under a cooperative agreement with the NSF on behalf of the Gemini partnership: the National Science Foundation (United States), the National Research Council (Canada), CONICYT (Chile), Ministerio de Ciencia, Tecnolog\'{i}a e Innovaci\'{o}n Productiva (Argentina), and Minist\'{e}rio da Ci\^{e}ncia, Tecnologia e Inova\c{c}\~{a}o (Brazil).

This paper makes use of the following ALMA data: ADS/JAO.ALMA\#2016.1.00126.S. ALMA is a partnership of ESO (representing its member states), NSF (USA) and NINS (Japan), together with NRC (Canada), NSC and ASIAA (Taiwan), and KASI (Republic of Korea), in cooperation with the Republic of Chile. The Joint ALMA Observatory is operated by ESO, AUI/NRAO and NAOJ. The National Radio Astronomy Observatory is a facility of the National Science Foundation operated under cooperative agreement by Associated Universities, Inc. This material is based upon work supported by the National Science Foundation Graduate Research Fellowship Program under grant No. DGE 1106400.

D.N. was funded in part by NSF AST-1715206 and HST AR-15043.0001.
\end{acknowledgements}



\end{document}